\documentclass[aps,prd,twocolumn,english]{revtex4-2}
\usepackage[T1]{fontenc}
\usepackage[utf8]{luainputenc}
\usepackage{amsmath,amsfonts,amsthm,bm}
\usepackage{braket}
\usepackage{amssymb}
\usepackage{amsthm}
\usepackage{graphicx}
\usepackage{esint}
\usepackage{babel}
\usepackage{longtable}
\usepackage{tabularx}
\usepackage{tikz}
\usetikzlibrary{arrows.meta,positioning}

\begin{document}

\title{Entanglement harvesting and curvature of entanglement: A modular operator approach}
	\author{Rupak Chatterjee}
 	\affiliation{
		Department of Applied Physics, New York University, 2 MetroTech Center, Brooklyn, NY, 11201}
	\email{Rupak.Chatterjee@nyu.edu}
    \date{\today}

\begin{abstract}
An operator-algebraic framework based on Tomita-Takesaki modular theory is used to study aspects of quantum entanglement via the application of the modular conjugation operator \( J \). The entanglement structure of quantum fields is studied through the protocol of entanglement
harvesting whereby quantum correlations evolve through the time evolution of qubit
detectors coupled to a Bosonic field. Modular conjugation operators are constructed for Unruh-Dewitt type qubits interacting with a scalar field such that initially unentangled
qubits become entangled. The entanglement harvested in this process is directly quantified by an expectation value involving \( J \) offering a physical application of this operator. The modular operator formalism is then extended to the Markovian open system dynamics of coupled qubits by expressing entanglement monotones as functionals of a state \(\rho\) and its modular reflection \( J\rho J \). The second derivative of such functionals with respect to an external coupling parameter, termed the curvature of entanglement, provides a natural measure of entanglement sensitivity. At points of modular self-duality, the curvature of entanglement coincides with the quantum Fisher information measure.
These results demonstrate that the modular conjugation operator $J$ captures both the harvesting of entanglement from quantum fields and the curvature of entanglement in coupled qubit dynamics providing parallel modular structures that connect these systems.
\end{abstract}

\maketitle

\section{Introduction}

The scientific significance of quantum entanglement lies in its ability to distinguish intrinsically quantum systems from classical ones. A related phenomenon, known as entanglement harvesting, occurs when two initially uncorrelated qubit detectors become entangled through local interactions with a background quantum field. More generally, entanglement harvesting refers to any process by which localized quantum systems extract preexisting entanglement from the vacuum or excited states of a quantum field. Quantum field theory predicts that the vacuum state contains quantum fluctuations, which can mediate entanglement between spatially separated qubits despite the absence of direct interaction. In this sense, the qubits locally interact with the field and harvest the entanglement present in the vacuum state. Within the framework of algebraic quantum field theory, it is well established that the ground state of a quantum field exhibits correlations between observables associated with spacelike-separated regions~\cite{Haag1}. Remarkably, these correlations can be strong enough to violate Bell-type inequalities~\cite{Summers1985}.

Furthermore, recent work has highlighted connections between quantum entanglement and certain geometric concepts, in particular the use of curvature to characterize the sensitivity of entanglement under parameter variations \cite{Saleem2025}. Within this context, the quantum Fisher information plays a dual role as both a precision bound in quantum estimation theory and a curvature measure for suitably defined entanglement functionals. In the present work, our focus is on the curvature of entanglement derived from operator-algebraic constructions involving the modular conjugation operator $J$. This perspective allows us to retain a clear emphasis on the modular structure while still drawing on curvature as a unifying quantitative descriptor.

This paper builds on a sequence of earlier works by the author ~\cite{Chatterjee1, Chatterjee2, Chatterjee3} exploring the connection between Tomita--Takesaki modular theory and quantum entanglement. In~\cite{Chatterjee1}, a modular operator expression for Wootter's concurrence \cite{Wootters1, Woot2001} was formulated in the setting of supersymmetric quantum mechanics, establishing a connection between entanglement and the antiunitary modular conjugation operator \( J \) for pure states. This framework was extended in~\cite{Chatterjee2} to causally closed regions in algebraic quantum field theory, where modular operators govern the entanglement structure of local algebras. In~\cite{Chatterjee3}, the modular formalism was applied to a physically realizable setting involving an Unruh--DeWitt detector locally interacting with a bosonic quantum field. The present work advances this program by extending the modular operator framework to the phenomenon of entanglement harvesting in quantum field theory. In this model, localized qubit detectors couple to a scalar field for finite durations, and despite remaining spacelike separated, they extract quantum correlations from the vacuum. Specifically, the expectation value of a suitably defined modular conjugation operator encodes the off-diagonal coherence induced by the field and reproduces the leading-order contribution to concurrence within perturbation theory.

We extend the use of the modular conjugation operator \( J \) to define quantum concurrence for mixed states by expressing it as a functional of both the state \( \rho \) and its modular reflection \( J \rho J \). The map \( \rho \mapsto J \rho J \) remains well defined for finite-dimensional density matrices and captures a form of modular symmetry. This enables us to construct entanglement monotones as operator functionals of the form \( f(\rho, J \rho J) \), which generalize the pure-state modular framework introduced in~\cite{Chatterjee1,Chatterjee2, Chatterjee3}. These functionals provide the foundation for our later analysis of concurrence and the curvature of entanglement.

The notion of entanglement curvature was recently introduced in~\cite{Saleem2025}, where the second derivative of concurrence with respect to an external control parameter \( g \) was shown to capture entanglement sensitivity in two-qubit systems. In this work, we generalize their framework by defining curvature for modular functionals of the form \( f(\rho, J \rho J) \), leading to the quantity
\[
K_E(g) := \frac{\partial^2}{\partial g^2} f(\rho(g), J \rho(g) J).
\]
We show that for a specific functional, this curvature coincides with the quantum Fisher information at points of modular self-duality, where \( \rho = J \rho J \). Modular self-duality serves as a fixed point of the modular reflection map, and at such points the symmetry of the functional amplifies its sensitivity to unitary perturbations.

In what follows, we work within the setting of type I von Neumann algebras corresponding to finite-dimensional Hilbert spaces of the effective detector models used here. In this context, density operators and tensor-product factorizations $\mathcal{H}_A \otimes \mathcal{H}_B$ are well defined ensuring that modular conjugation can be represented by a fixed antiunitary operator acting on a separable Hilbert space and permitting explicit constructions of functionals of the form $f(\rho, J\rho J)$.

To illustrate this approach concretely, we analyze a dissipative cavity QED system. The exact solution for the time-evolved density matrix allows us to compute the concurrence and its curvature analytically. We verify that the modular overlap structure encoded in \( f(\rho, J \rho J) \) faithfully tracks the behavior of concurrence and quantum Fisher information across time and coupling parameter domains. At points of maximal entanglement, the modular curvature equals the quantum Fisher information, and the modular reflection symmetry exhibits alignment with the structure of the system's dynamical evolution. This suggests that Tomita–Takesaki theory not only encodes algebraic duality, but also captures the local information geometry of quantum states. The connection to quantum estimation theory becomes natural whereby the quantum Fisher information measure determines the optimal precision with which parameters can be inferred from a quantum system while simultaneously emerging as a curvature measure that quantifies the rate of entanglement deformation.

\bigskip

The remainder of this paper is organized as follows. In Sec.~II, we review some elements of Tomita--Takesaki modular theory and define the modular conjugation operator and modular reflection that underpin our framework. In Sec.~III, we apply this structure to entanglement harvesting using Unruh--DeWitt detectors interacting with a scalar quantum field, and show how a modular expectation value quantifies the harvested entanglement. In Sec.~IV, we reformulate Wootters’ concurrence for mixed states as a modular functional \( f(\rho, J \rho J) \) in order to reinterpret entanglement through modular reflection. In Sec.~V, we introduce the concept of modular curvature of entanglement as the second derivative of such functionals with respect to a coupling parameter.  In Sec.~V-A, we revisit the coupled qubit model and recover the curvature of entanglement behavior previously studied in~\cite{Saleem2025}, now interpreted through our operator-algebraic framework. We show that this curvature coincides with quantum Fisher information at points of modular self-duality. In  Sec.~V-B, we discuss the second-order expansion of relative entropy, and the emergence of quantum Fisher information as a curvature measure tied to modular self-duality.
Finally, in Sec.~VI, we outline future directions, including a broader class of modular functionals and conjectured monotone Riemannian metrics defined via the modular conjugation operator, thereby embedding entanglement curvature into a unified operator-algebraic framework. Additional technical derivations are included in the Appendix.

\section{Tomita--Takesaki Modular Operators}

We begin by reviewing the essential elements of Tomita--Takesaki modular theory and defining the key operators that will underlie our formulation of entanglement. Let \( \mathcal{B}(\mathcal{H}) \) denote the algebra of bounded linear operators on a Hilbert space \( \mathcal{H} \). Consider a \( \bm{C}^* \)-subalgebra \( \mathcal{A} \subset \mathcal{B}(\mathcal{H}) \), and define its \emph{commutant} \( \mathcal{A}' \subset \mathcal{B}(\mathcal{H}) \) as the set of all bounded operators commuting with every element of \( \mathcal{A} \), i.e.,
\[
\mathcal{A}' := \{ B \in \mathcal{B}(\mathcal{H}) \mid [A, B] = 0 \text{ for all } A \in \mathcal{A} \}.
\]
If the double commutant satisfies \( \mathcal{A}'' = \mathcal{A} \), then \( \mathcal{A} \) (and hence \( \mathcal{A}' \)) is a von Neumann algebra.

The duality between \( \mathcal{A} \) and \( \mathcal{A}' \) is made explicit through the Tomita--Takesaki modular framework. Let \( |\Omega\rangle \in \mathcal{H} \) be a \emph{cyclic and separating vector} for \( \mathcal{A} \), meaning that the set \( \{ A|\Omega\rangle \mid A \in \mathcal{A} \} \) is dense in \( \mathcal{H} \), and that \( A|\Omega\rangle = 0 \Rightarrow A = 0 \) for all \( A \in \mathcal{A} \).
Tomita--Takesaki theory asserts that there exists a densely defined, anti-linear operator \( S: \mathcal{H} \to \mathcal{H} \) such that
\[
S A |\Omega\rangle = A^* |\Omega\rangle, \quad \forall A \in \mathcal{A},
\]
where \( A^* \) denotes the \( * \)-involution from the \( \bm{C}^* \)-algebra structure. This operator admits a polar decomposition,
\[
S = J \Delta^{1/2} = \Delta^{-1/2} J,
\]
where \( \Delta = S^* S \) is the \emph{modular operator} and \( J \) is the \emph{modular conjugation operator}. These satisfy the defining identities:
\begin{equation}
    \begin{array}{cc}
        J \Delta^{1/2} J = \Delta^{-1/2}, \;\;\;
        J^2 = I, \;\;\; J^* = J, \\\\
        J |\Omega\rangle = |\Omega\rangle, \;\;\;
        J \mathcal{A} J = \mathcal{A}'.
    \end{array}
\end{equation}

The modular conjugation operator \( J \) implements a duality between \( \mathcal{A} \) and its commutant \( \mathcal{A}' \), while the modular operator \( \Delta \) generates a one-parameter group of automorphisms on \( \mathcal{A} \), known as the \emph{modular flow}:
\begin{equation}
\sigma_t(A) := \Delta^{it} A \Delta^{-it}, \quad t \in \mathbb{R}.
\end{equation}
This flow leaves the algebra \( \mathcal{A} \) invariant and encodes a canonical dynamics determined entirely by the algebra and the state \( |\Omega\rangle \). Further operator-theoretic details are provided in Appendix A.

We interpret the modular conjugation operator \( J \) as implementing a \emph{reflection symmetry} across the algebraic bipartition defined by \( \mathcal{A} \) and its commutant \( \mathcal{A}' \), exchanging \( \mathcal{A} \leftrightarrow \mathcal{A}' \). Given a density matrix \( \rho \), we define its \emph{modular reflection} as
\begin{equation}
\rho_J := J \rho J.
\end{equation}
This operator serves as a canonical dual to \( \rho \). We say that a state \( \rho \) exhibits \emph{modular self-duality} if
\begin{equation}
\rho = J \rho J,
\end{equation}
i.e., the state \( \rho \) is invariant under modular reflection. This condition defines a fixed point under the algebraic reflection symmetry and, as we will show, corresponds to maximal entanglement in certain physical models. When \( \rho \ne J \rho J \), the state breaks modular self-duality and thereby distinguishes between the algebra \( \mathcal{A} \) and its commutant \( \mathcal{A}' \).

In what follows, we exploit the modular reflection structure \( \rho \mapsto J \rho J \) to construct entanglement monotones and information-theoretic quantities directly from operator-algebraic data. In particular, we analyze functionals of the form \( f(\rho, J \rho J) \) and interpret their second derivatives with respect to external parameters as measures of entanglement curvature. This framework allows us to extend our earlier work using modular tools~\cite{Chatterjee1,Chatterjee2,Chatterjee3} to mixed states. A finite-dimensional realization of modular reflection for two-qubit systems is worked out explicitly in Appendix~B.

For the Bosonic field theory model with Unruh-DeWitt qubit detectors studied below, we will need some further elaboration following \cite{ summers1987I, summers1987II, Eckmann1973}.  Let $f \in C_0^{\infty}(\mathcal{O})$ be smooth complex functions with compact support in an open region of Minkowski spacetime $\mathcal{O}$ and denote the space of such test functions as $S(\mathcal{O})$. Define the complement of $S(\mathcal{O})$, denoted by 
$S(\mathcal{O}')$, as the space of test functions with vanishing functional inner product with those in  $S(\mathcal{O})$, 
\begin{equation}
\begin{array}{c}
 \braket{f,g} = \displaystyle\int f(x) G(x-y) g(y) d^4x d^4y =0,\\\\
 f \in S(\mathcal{O}), \; g \in S(\mathcal{O}').
 \end{array}
\end{equation}
where $G(x-y)$ is a propagator such as a Pauli-Jordan Green's function.
This condition ensures that the smeared field operators commute,
\begin{equation}
 \; [\phi(f),\phi(g)]=0, \;\;\;
 \forall f \in S(\mathcal{O}) \;\text{and}\; \forall \; g \in S(\mathcal{O}')
\end{equation}
expressing microcausality in the language of test function supports (see \cite{Chatterjee3} for further elaboration).

Now consider a von Neumann algebra $\mathcal{A}(S(\mathcal{O}))$ of Weyl style operators $A=e^{i\phi(f)}$ with $f \in S(\mathcal{O})$. Using Haag duality \cite{Haag1} for free Bose fields, one can state that $\mathcal{A}'(S(\mathcal{O})) = \mathcal{A}(S(\mathcal{O}'))$ such that
the commutant $\mathcal{A}'$ is seen to be equivalent to the algebra $\mathcal{A}$ defined on the complement space of test functions. Eckmann and Osterwalder~\cite{Eckmann1973} showed that the modular conjugation operator \( J \) acts on Weyl operators  $Je^{i\phi(f)}J=e^{-i\phi(jf)}$ where $f \in S(\mathcal{O})$ with $jf \in S(\mathcal{O}')$. This construction lifts the Tomita-Takesaki conjugation operator to the level of test functions via the map \( j: S(\mathcal{O}) \to S(\mathcal{O}') \). Finally, for 
$f \in S(\mathcal{O}), \; g \in S(\mathcal{O}')$, \cite{summers1987II, Fabritiis2023} derive the following vacuum state expectation value for products of Weyl operators,
\begin{equation}
 \bra{0}e^{i\phi(f)}e^{i\phi(g)}\ket{0} = \bra{0}e^{i\phi(f+g)}\ket{0} = e^{-\frac{1}{2}||f+g||^2} ,\label{WeylExpectation}
\end{equation}
where $||h||=\braket{h,h}$.
We will apply this framework to a physical model where qubit entanglement is generated dynamically by coupling to a scalar quantum field. This will allow us to demonstrate how modular conjugation governs both entanglement harvesting and the structure of concurrence in a spacetime-localized quantum field setting. 

\section{Entanglement Harvesting and Concurrence for Unruh-Dewitt Type Qubits in a Scalar Quantum Field.}

We begin by applying the Tomita–Takesaki modular framework to a physical setting where entanglement arises dynamically.
The Unruh-DeWitt particle detector model \cite{Unruh1976, DeWitt1979, Tjoa2023} is a quantum mechanical system where a qubit (the 'detector') interacts with a Bosonic quantum field. The detector coupled to the quantum field allows it to absorb or emit the quanta of the scalar field leading to transitions between its energy levels. The Unruh-DeWitt detector was first used to study the Unruh effect where an observer accelerating through empty space will detect a thermal bath of particles even though an inertial observer would see none. Let the Hilbert spaces of the detector and the field be denoted as \( \mathcal{H}_D \) and \( \mathcal{H}_F \) with the total Hilbert space given by \( \mathcal{H} = \mathcal{H}_D \otimes \mathcal{H}_F \).
The detector Hamiltonian \( H_D \) is
\begin{equation}
H_D = \frac{1}{2}  \omega (\sigma^z+I_D) \otimes I_F,
\end{equation}
where \( \sigma^z \) is the Pauli matrix acting on \( \mathcal{H}_D \), and \( I_F \) is the identity operator on \( \mathcal{H}_F \).
The field Hamiltonian \( H_F \) is
\begin{equation}
H_F = I_D \otimes \int d^3x \left( \frac{1}{2} \pi^2 + \frac{1}{2} (\nabla \phi)^2 + \frac{1}{2} m^2 \phi^2 \right),
\end{equation}
where \( I_D \) is the identity operator on \( \mathcal{H}_D \).
The coupling term \( H_I \) is chosen as
\begin{equation}
\begin{array}{c}
H_I (\tau)=  \lambda \chi(x(\tau))\left[\sigma_{+} e^{i\omega \tau} + \sigma_{-} e^{-i\omega \tau}\right] \otimes \phi(x(\tau)),
\end{array}
\end{equation}
where $\lambda$ is the interaction strength, and \( \sigma_+ \) and \( \sigma_- \) are the raising and lowering operators acting on \( \mathcal{H}_D \)
\begin{equation}
    \begin{array}{c}
   \sigma_+  = \ket{1}\bra{0}      \\\\
    \sigma_-  =  \ket{0}\bra{1}
    \end{array}
\end{equation}
\( \phi(x(\tau)) \) is the field operator acting on \( \mathcal{H}_F \) evaluated along the detector's trajectory with $\tau$ being the proper time. The function \( \chi(x(\tau)) \) is a smooth, compactly supported switching function that controls both the location and duration of the interaction between the qubit and the scalar field along the detector’s worldline. The width of \( \chi \) determines the effective interaction time.
 Unitary evolution in the interaction picture is given by the time-ordered expression up to $\mathcal{O}(\lambda^3) $
\begin{equation}
\begin{array}{c}
 U=\mathcal{T}\exp\left[-i\displaystyle\int d\tau H_I (\tau) \right]
 = I -i\displaystyle\int d\tau H_I (\tau) \\\\ +\dfrac{(-i)^2}{2}\displaystyle\int\int d\tau d\tau' \mathcal{T}\left\{H_I (\tau)H_I (\tau')\right\} +\mathcal{O}(\lambda^3).
 \end{array}
 \end{equation}
\( \mathcal{T} \) denotes time-ordering along the detector’s proper time \( \tau \) ensuring that operators are ordered chronologically along the detector's worldline,
 \begin{equation}
   \mathcal{T}A(\tau)B(\tau') = \theta[\tau - \tau'] A(\tau)B(\tau') +\theta[\tau' - \tau]B(\tau')A(\tau)
 \end{equation}
 where $\theta[\tau - \tau']$ is the Heaviside step function imposing causal ordering.

To probe the entanglement structure of a scalar field, consider two Unruh-DeWitt detectors labeled by $A$ and $B$ following trajectories $x_A(\tau_A))$ and $x_B(\tau_B))$. These detectors couple to the real scalar field $\phi(x(\tau))$ via the interaction Hamiltonians $H^A_I (\tau_A)$ and $H^B_I (\tau_B)$ such that the Hilbert space structure is $\mathbb{C}_A^2 \otimes \mathbb{C}_B^2 \otimes \mathcal{H}_F$. Working in the interaction picture, the unitary time evolution operator up to \( \mathcal{O}(\lambda^3) \) is explicitly given by~\cite{Smith2019}
\begin{equation}
\begin{array}{c}
 U= I -i\displaystyle\int dt \dfrac{d\tau_A }{dt}H^A_I [\tau_A (t)] -i\displaystyle\int dt \dfrac{d\tau_B }{dt}H^B_I [\tau_B(t)]\\\\ +\dfrac{(-i)^2}{2}\displaystyle\int\int dt dt'   \dfrac{d\tau_A }{dt} \dfrac{d\tau_A }{dt'} \mathcal{T}\left\{H^A_I [\tau_A (t)]H^A_I [\tau_A (t')]\right\} \\\\ +\dfrac{(-i)^2}{2}\displaystyle\int\int dt dt'   \dfrac{d\tau_B }{dt} \dfrac{d\tau_B }{dt'} \mathcal{T}\left\{H^B_I [\tau_B (t)]H^B_I [\tau_B (t')]\right\} \\\\ +\dfrac{(-i)^2}{2}\displaystyle\int\int dt dt'   \dfrac{d\tau_A }{dt} \dfrac{d\tau_B }{dt'} \mathcal{T}\left\{H^A_I [\tau_A (t)]H^B_I [\tau_B (t')]\right\} \\\\ +\dfrac{(-i)^2}{2}\displaystyle\int\int dt dt'   \dfrac{d\tau_B }{dt} \dfrac{d\tau_A }{dt'} \mathcal{T}\left\{H^B_I [\tau_B (t)]H^A_I [\tau_A (t')]\right\} 
 \\\\+\mathcal{O}(\lambda^3)
 \end{array}
 \end{equation}
 Here, the Jacobian factors \( \frac{d\tau}{dt} \) arise from the reparameterization of the detector's proper time \( \tau \) in terms of coordinate time.
 
 Using this expansion, perturbation theory follows for unitary time evolution $\ket{\Psi_f}=U\ket{\Psi_i}=\sum_n \lambda^n \ket{\Psi_f}^{(n)}$ where, for instance starting from an initial separable state,
 \begin{equation}
    \ket{\Psi_f}^{(0)}=\ket{0}_A \otimes \ket{0}_B \otimes \ket{0}_{\phi},
 \end{equation}
 one has the first order correction given by
 \begin{equation}
 \begin{array}{c}
    \ket{\Psi_f}^{(1)}=\\\\
    -i \displaystyle\int dt \chi(x_A(\tau_A))\dfrac{d\tau_A }{dt}e^{i \omega_A \tau_A(t)}\ket{1}_A \otimes \ket{0}_B \otimes \phi_A(t)\ket{0}_{\phi} \\\\-i \displaystyle\int dt \chi(x_B(\tau_B))\dfrac{d\tau_B }{dt}e^{i \omega_B \tau_B(t)}\ket{0}_A \otimes \ket{1}_B \otimes \phi_B(t)\ket{0}_{\phi}
\end{array}
\end{equation}
(see \cite{Smith2019} for higher order terms).

As shown in \cite{Chatterjee1, Chatterjee2, Chatterjee3}, the modular conjugation operator \( J \), defined via Tomita--Takesaki theory, may serve as a basis for constructing entanglement measures in pure quantum states.
 We define the modular conjugation operator \( J_{AB} : \mathbb{C}_A^2 \otimes \mathbb{C}_B^2 \rightarrow \mathbb{C}_A^2 \otimes \mathbb{C}_B^2 \) acting anti-unitarily on tensor product basis states as (see Appendix~B)

\begin{equation}
J_{AB}\left[
\begin{pmatrix}
\alpha \\
\beta
\end{pmatrix} \otimes
\begin{pmatrix}
\gamma \\
\delta
\end{pmatrix}  \right]
=
\begin{pmatrix}
\bar{\delta} \\
\bar{\gamma}
\end{pmatrix} \otimes
\begin{pmatrix}
\bar{\beta} \\
\bar{\alpha}
\end{pmatrix} .
\label{eq:JAB}
\end{equation}

Calculating the expectation value of this operator $\braket{\Psi_f |J_{AB} \otimes I_F|\Psi_f}$ using the time-evolved entangled state $\Psi_f$ including second order terms $\lambda^2$ produces (see Appendix C for a detailed derivation)
\begin{equation}
\begin{array}{c}
\left|\left(\bra{\Psi_f}^{(0)}+\lambda \bra{\Psi_f}^{(1)} +\lambda^2 \bra{\Psi_f}^{(2)}\right)\right. \\\\
\times (J_{AB} \otimes I_F) \\\\   \times\left.\left(\ket{\Psi_f}^{(0)}+\lambda \ket{\Psi_f}^{(1)}+\lambda^2 \ket{\Psi_f}^{(2)}\right)\right| \\\\=
2|X|+\mathcal{O}(\lambda^4)
\end{array} \label{keyResult}
\end{equation}
where
\begin{equation}
\begin{array}{c}
X = \\-\lambda^2 \displaystyle \int d\tau_A  \int d\tau_B \chi\bigl(x_A(\tau_A)\bigr)\chi\bigl(x_B(\tau_B)\bigr)e^{-i(\omega_A \tau_A  + \omega_B \tau_B)} \\\\ \times 
\biggl[\theta[t_B(\tau_B)-t_A(\tau_A)]\braket{0|\phi(x_A(\tau_A))\phi(x_B(\tau_B))|0}\biggr.\\\\+\biggl.(A \leftrightarrow B)\biggr]\\\\
\end{array}
\end{equation}
(\cite{Smith2019}).

How is this related to more conventional measures such as concurrence? Denoting $P_D$ as the probability that a qubit detector $D=\{A,B\}$ has transitioned to its excited
\begin{equation}
\begin{array}{c}
P_D = \lambda^2 \displaystyle \int d\tau  \int d\tau' \chi\bigl(x_D(\tau)\bigr)\chi\bigl(x_D(\tau')\bigr)\\\\
\times  e^{-i\omega_D(\tau  - \tau')} 
\braket{0|\phi(x_D(\tau))\phi(x_D(\tau'))|0},
\end{array}
\end{equation}
quantum concurrence is given by \cite{Smith2019}
\begin{equation}
    C=2\max\biggl[0, |X| - \sqrt{P_A P_B} \biggr]+\mathcal{O}(\lambda^4)
\end{equation}
The final state \( \ket{\Psi_f} \) exhibits maximal entanglement when the coherence term \( |X| \) dominates over the product of individual excitation probabilities \( \sqrt{P_A P_B} \).

 The term \( 2|X| \), arising entirely from field-induced off-diagonal coherence, is precisely the modular expectation value \( \braket{\Psi_f | J_{AB} \otimes I_F | \Psi_f} \).
This quantity captures the entanglement generated between the detectors due to spacelike correlations in the quantum field, as encoded by the Tomita--Takesaki modular conjugation operator.
Although \(J_{AB}\) is antiunitary and therefore not a Hermitian observable, the overlap \( \braket{\Psi_f | J_{AB} \otimes I_F | \Psi_f} \) can be interpreted operationally as an interference amplitude between the final state and its modularly reflected counterpart.
Its modulus, \(2|X|\), quantifies the visibility of coherence between excitation pathways that are exchanged under detector interchange and complex conjugation.
Interferometric and weak-measurement protocols have been shown to directly access such overlap amplitudes and coherences, without requiring full state tomography~\cite{ekert2002,flammia2011,dressel2014,lundeen2011}.
In this sense, the quantity \(2|X|\) plays the same operational role as measurable coherence and fidelity estimators in resource-theoretic formulations of quantum coherence~\cite{baumgratz2014}.
Thus, while \(J_{AB}\) itself is not directly measurable, the interference pattern it defines possesses clear operational meaning and coincides with the concurrence’s coherence term. These same interference visibilities appear in recent optical entanglement-harvesting experiments and photonic state-overlap measurements suggesting possible realizations of the modular overlap in cavity-QED or circuit-QED platforms.

In contrast, the subtraction term \( \sqrt{P_A P_B} \) involves the populations of the excited states and is not accessible via the antiunitary operator \( J_{AB} \), which acts only on coherence terms and leaves diagonal elements invariant. Specifically, \( J_{AB} \) swaps and conjugates qubit components in a way that does not affect the populations of individual detector states. The excitation probabilities, by contrast, are encoded in the diagonal matrix elements:
\begin{equation}
\begin{array}{c}
P_A = \langle \Psi_f | \Pi_A \otimes I_B \otimes I_\phi | \Psi_f \rangle\,, \\\\
P_B = \langle \Psi_f | I_A \otimes \Pi_B \otimes I_\phi | \Psi_f \rangle\,,
\end{array}
\end{equation}
where \( \Pi_A = |1\rangle\langle 1|_A \) and \( \Pi_B = |1\rangle\langle 1|_B \) are projection operators onto the excited detector states. Using the norms of these projected states, we obtain
\begin{equation}
\| \Pi_A |\Psi_f\rangle \| = \sqrt{P_A}\,, \qquad \| \Pi_B |\Psi_f\rangle \| = \sqrt{P_B}\,,
\end{equation}
and thus the subtraction term appearing in the concurrence expression becomes
\begin{equation}
\sqrt{P_A P_B} = \| \Pi_A |\Psi_f\rangle \| \cdot \| \Pi_B |\Psi_f\rangle \|\,.
\end{equation}

We note that the modular conjugation operator \( J_{AB} \) defined in Eq.~(\ref{eq:JAB}), which acts by conjugating coefficients and reversing tensor factors, differs in appearance from the standard spin-flip operator commonly used in quantum information theory (QIT), \( J_{\mathrm{QIT}} = (\sigma_y \otimes \sigma_y) K \), where $K$ denotes complex conjugation in the computational basis. As shown in Appendix~B, these operators are unitarily equivalent up to a local SWAP and global phase. This ensures that the expectation values and symmetry properties computed here remain consistent with conventional treatments based on spin-flip conjugation.

The above results are somewhat reminiscent of earlier results by Summers and Werner~\cite{summers1987I, summers1987II} who demonstrated that the vacuum of a relativistic free quantum field exhibits entanglement between spacelike separated regions. Moreover, they showed that these vacuum correlations can violate Bell inequalities — a manifestation of nonlocality analogous to the Reeh–Schlieder theorem.

\section{Modular Formulation of Concurrence for Mixed States}

For two qubit mixed states described by a density matrix $\rho$, the traditional concurrence is defined via the standard Wootters procedure \cite{Wootters1, Woot2001}. One first constructs the spin-flipped operator
\begin{equation}
\tilde{\rho} = (\sigma_y \otimes \sigma_y)\, \rho^*\, (\sigma_y \otimes \sigma_y),
\end{equation}
then forms the Hermitian matrix
\begin{equation}
R = \rho\, \tilde{\rho},
\end{equation}
and finally defines the concurrence as
\begin{equation}
C(\rho) = \max\left\{ 0,\ \sqrt{\lambda_1} - \sqrt{\lambda_2} - \sqrt{\lambda_3} - \sqrt{\lambda_4} \right\},
\end{equation}
where $\lambda_i$ are the eigenvalues of $R$ arranged in decreasing order. This definition, though operationally effective, obscures the algebraic and modular structure underlying entanglement.

To extend Tomita--Takesaki theory to mixed states, we employ the GNS construction associated with the state $\rho$.
Let \( \rho \) be a density operator acting on \( \mathcal{H}_A \otimes \mathcal{H}_B \), and let \( J \) denote the Tomita--Takesaki modular conjugation operator with respect to a von Neumann algebra \( \mathcal{A} \subset \mathcal{B}(\mathcal{H}_A \otimes \mathcal{H}_B) \) and a cyclic separating vector \( |\Omega\rangle \) representing \( \rho \).
If \( \rho \) is a pure state and is maximally entangled, its concurrence satisfies $C(\rho) = 1$.
More generally, if \( \rho = J \rho J \) but \( \rho \) is mixed, then the concurrence is bounded above by unity but not necessarily at maximal entanglement.
This establishes that exact modular self-duality is a necessary condition for maximal entanglement in pure states and positions concurrence as a functional that quantifies its violation. 

It is important to distinguish between the pure- and mixed-state cases when discussing modular self-duality. 
For \textit{pure states}, modular self-duality \( \rho = J\rho J \) is both a \textit{necessary and sufficient} condition for maximal entanglement, since it coincides with equal Schmidt coefficients and \( C(\rho) = 1 \).

However, for \textit{mixed states}, modular self-duality remains \textit{necessary but not sufficient} for maximal entanglement. In particular, any state that is diagonal in a basis symmetric under \(J\) (such as the maximally mixed state) trivially satisfies this equality while possessing zero concurrence. Maximal entanglement arises only when self-duality holds together with additional spectral and support conditions—namely, when the support of \(\rho\) lies within the Bell subspace generated by the eigenvectors of \(R = \rho J\rho J\) with non-zero off-diagonal coherence. Equivalently, for rank-one (pure) or rank-two states with equal Schmidt coefficients, modular self-duality coincides with \(C(\rho) = 1\). This distinction clarifies that \(J\rho J = \rho\) characterizes the fixed-point symmetry, while concurrence quantifies the degree of its violation.

The physical content of this construction is reinforced by its connection to Bell inequalities. In particular, the algebraic concurrence defined above for two qubit states satisfies the maximal Bell-CHSH inequality violation,
\begin{equation}
|\langle B \rangle|_{\text{max}} = 2 \sqrt{1 + C(\rho)^2},
\end{equation}
confirming its consistency with operational measures of quantum nonlocality \cite{Chatterjee3}.

To further clarify the construction, one can compute $C(\rho)$ explicitly for a mixed state, using the QIT modular conjugation operator (see Appendix B for further explanation)
\begin{equation}
J_{\mathrm{QIT}} = (\sigma_y \otimes \sigma_y) K,
\end{equation}
where $K$ denotes complex conjugation in the computational basis. Then the modular reflection of the mixed state is given by 
\begin{equation}
\rho_J =J_{\mathrm{QIT}} \; \rho \; J_{\mathrm{QIT}} = (\sigma_y \otimes \sigma_y)\, \rho^*\, (\sigma_y \otimes \sigma_y),
\end{equation}
so that
\begin{equation}
R = \rho\, \rho_J = \rho\, \bigl( J_{\mathrm{QIT}} \; \rho \; J_{\mathrm{QIT}}\bigr).
\end{equation}
This form reveals that entanglement is encoded in the interaction between $\rho$ and its modular reflection $J \rho J$ with $R$ quantifying the deviation from self-duality.  Our contribution is to reinterpret this algebraic structure as a modular reflection within the framework of Tomita–Takesaki theory, thereby embedding concurrence into a broader algebraic setting. We will apply the above methodology to a coupled qubit model below.

In the finite-dimensional setting considered here, the modular conjugation operator is defined within a fixed representation of the type I von Neumann algebra 
\(\mathcal{A}\subset\mathcal{B}(\mathcal{H}_A\otimes\mathcal{H}_B)\).
The operator \(J_{\mathrm{QIT}}=(\sigma_y\otimes\sigma_y)K\) acts as an antiunitary reflection that exchanges the two subsystems and performs complex conjugation in the computational basis.
As detailed in Appendix~B, any other representation obtained by local unitaries \(U_A\otimes U_B\) yields an equivalent conjugation 
\(J'=(U_A\otimes U_B)J_{\mathrm{QIT}}(U_A^\dagger\otimes U_B^\dagger)\),
so the modular structure is representation dependent but algebraically fixed.
Once a computational representation is chosen, \(J\) therefore remains invariant under all physical operations within that representation.
This representational invariance underlies the subsequent analysis, where the same modular conjugation is used to define curvature functionals of the density operator.

\section{Curvature of Entanglement}
We now investigate how curvature of entanglement arises from operator-algebraic symmetry, rather than relying solely on heuristic inversion or spectral constructions.
Let \( \rho(g,t) \) be a smooth one-parameter family of density operators, where \( g \) is a real-valued coupling parameter.
 Following \cite{Saleem2025}, the curvature of entanglement was defined as the second partial derivative of concurrence with respect to \( g \),
$\mathcal{K}_E(g,t) = \dfrac{\partial^2 C(\rho(g,t))}{\partial g^2}.$
This quantity measures the sensitivity of concurrence to infinitesimal variations in the coupling parameter \( g \).
Its connection to quantum estimation theory arises from the fact that the quantum Fisher information governs the local distinguishability of nearby states under deformations in \( g \)—precisely the variation encoded in the second derivative of an entanglement monotone.

In our framework, given that \( C(\rho) = f(\rho, J \rho J) \) where \( f \) is a scalar-valued entanglement functional that is Fr\'{e}chet-differentiable in both operator arguments~\cite{ReedSimon,KadisonRingrose}, we can compute its second derivative with respect to the coupling parameter \( g \) as follows.
 The partial derivatives \( \frac{\partial f}{\partial \rho} \) and \( \frac{\partial f}{\partial (J \rho J)} \) are operator-valued elements in \( \mathcal{B}(\mathcal{H}) \), and their pairing with variations such as \( \frac{\partial^2 \rho}{\partial g^2} \) is naturally expressed using using the Hilbert--Schmidt inner product \cite{SimonBarry,Bhatia1997},
\begin{equation}
\langle A, B \rangle = \mathrm{Tr}(A^\dagger B) .
\end{equation}
This yields the following formal second-order expansion
\begin{equation}
\frac{\partial^2 C}{\partial g^2}
= \left\langle \frac{\partial f}{\partial \rho}, \frac{\partial^2 \rho}{\partial g^2} \right\rangle
+ \left\langle \frac{\partial f}{\partial (J \rho J)}, \frac{\partial^2 (J \rho J)}{\partial g^2} \right\rangle
+ \text{cross terms},
\end{equation}
where the inner products can be explicitly written as traces.

As established in Sec.~IV, \(J\) acts within a fixed computational representation of the type~I algebra and is therefore treated as constant when evaluating parameter derivatives.
Since the modular conjugation operator \(J\) is antiunitary and representation-dependent but not dynamically dependent on the coupling parameter \(g\), we have
\begin{equation}
\frac{\partial (J \rho J)}{\partial g} = J \!\left( \frac{\partial \rho}{\partial g} \right)\! J, 
\qquad
\frac{\partial^2 (J \rho J)}{\partial g^2} = J \!\left( \frac{\partial^2 \rho}{\partial g^2} \right)\! J.
\end{equation}
As a result, the curvature of entanglement assumes the following trace form
\begin{equation}
\frac{\partial^2 C}{\partial g^2} = \mathrm{Tr}\left[ \left( \frac{\partial f}{\partial \rho} + J \left( \frac{\partial f}{\partial (J \rho J)} \right) J \right) \cdot \frac{\partial^2 \rho}{\partial g^2} \right] + \cdots 
\end{equation}

We assume that $\rho(g)$ defines a twice continuously Fr\'{e}chet-differentiable trajectory in the Banach manifold of trace-class density operators with unit trace denoted as  $T_1(\mathcal{H})$ \cite{KadisonRingrose,SimonBarry}. Additionally, the entanglement functional $f(\rho, J\rho J)$ is assumed to be Fr\'{e}chet-differentiable in both operator arguments on a suitable convex subset of $T_1(\mathcal{H}) \times T_1(\mathcal{H})$. Such differentiability conditions are typically satisfied when $f$ is defined through operator monotone functions or spectral calculus of positive operators. These assumptions ensure that second derivatives such as $\partial^2\rho/\partial g^2$ and the associated operator-valued derivatives of $f$ remain bounded operators in $B(\mathcal{H})$ making them well-defined and ensuring trace-class integrability. Consequently, the Hilbert--Schmidt inner product $\langle A, B\rangle = \mathrm{Tr}(A^\dagger B)$ used throughout this work is justified.

When the quantum state depends smoothly on an external parameter \( g \), such as a coupling constant or interaction strength, it defines a differentiable trajectory
\begin{equation}
g \mapsto \rho(g) \in \mathcal{D}(\mathcal{H}),
\end{equation}
where \( \mathcal{D}(\mathcal{H}) \) denotes the space of density operators on the Hilbert space \( \mathcal{H} \). This trajectory describes the flow of the quantum state through the space of physical states as the parameter \( g \) varies. The first derivative \( \frac{\partial \rho}{\partial g} \) defines a tangent vector to this path in operator space, while the second derivative of a scalar functional—such as \( f(\rho(g), J \rho(g) J) \)—captures its curvature in the sense of information geometry.

Within this framework, we define the \emph{modular curvature of entanglement} as
\begin{equation}
\mathcal{K}_E(g) := \frac{\partial^2}{\partial g^2} f(\rho(g), J \rho(g) J),
\end{equation}
where the entanglement monotone $C(\rho)$ is now represented through the modular functional $f(\rho,J\rho J)$.
This quantifies the second-order sensitivity of entanglement to perturbations in the external parameter \( g \). This expression measures how the entanglement functional changes under infinitesimal variations in the state \( \rho(g) \) and its modular image \( J \rho(g) J \), thereby probing the underlying symmetry encoded by modular conjugation.
The modular operator \( J \) serves as a fixed, antiunitary reference that encodes the dual structure of entanglement. The curvature \( \mathcal{K}_E(g) \) quantifies the deviation of \( \rho(g) \) from exact modular self-duality as \( g \) varies, thus offering a geometric measure of entanglement sensitivity.
Maximal entanglement corresponds to modular self-duality \( \rho = J \rho J \) while degraded entanglement reflects increasing deviation from this fixed point. 

While the above general trace-based formula for the curvature of entanglement is not explicitly employed in the detailed model analyses presented subsequently, it nonetheless serves a foundational role. Specifically, this trace formulation provides a rigorous operator-algebraic underpinning and theoretical justification for the eigenvalue-based curvature calculations used below. In particular, concurrence in the example below is treated as a spectral functional of the pair $(\rho, J\rho J)$, depending only on the spectrum of the modular product $R = \rho(J\rho J)$. The trace formulation ensures that such spectral functionals fit naturally into the modular curvature framework, even when computed directly through their eigenvalue expressions.

\subsection{Coupled Qubits}

To illustrate the operational meaning of modular curvature, we analyze the dynamics of a two-qubit system governed by a coherent flip-flop interaction first studied with respect to entanglement curvature in~\cite{Saleem2025}. This model is inspired by a Jaynes-Cummings qubit-field Hamiltonian where an effective two-qubit interaction is obtained by adiabatic elimination of the field mode retaining only the excitation-exchange terms. This interaction captures coherent energy transfer between two-level systems and is experimentally realized in platforms such as cavity QED (with two atoms exchanging virtual photons), trapped ion systems, and superconducting qubits~\cite{Haroche,Blais}.
The system consists of two interacting qubits with the effective Hamiltonian
\begin{equation}
H_{\text{int}} = g\big(\sigma_+ \otimes \sigma_- + \sigma_- \otimes \sigma_+\big),
\end{equation}
where \(g\) is the dipole–dipole coupling strength, and \(\sigma_{\pm}\) are the standard Pauli raising and lowering operators. This Hamiltonian can be derived from a Jaynes--Cummings qubit--field model in the dispersive regime by eliminating the field mode, leaving only a direct qubit--qubit excitation exchange.
The open-system dynamics are described by a Markovian Lindblad master equation:
\begin{equation}
\begin{array}{c}
\dfrac{d\rho}{dt} = -i[H_{\text{int}}, \rho]
\\[6pt]
+ \dfrac{\kappa}{2}\displaystyle\sum_{j=1}^{2}\left(2\,\sigma_-^{(j)} \rho \sigma_+^{(j)} - \sigma_+^{(j)}\sigma_-^{(j)} \rho - \rho\, \sigma_+^{(j)}\sigma_-^{(j)}\right),
\end{array}
\end{equation}
where \(\kappa\) denotes the uniform decay rate of the qubits to their ground states.The superscripts on the Pauli ladder operators indicate which qubit the operator acts on: 
$\sigma_\pm^{(1)} = \sigma_\pm \otimes I$ acts only on the first qubit and leaves the second qubit unchanged, while 
$\sigma_\pm^{(2)} = I \otimes \sigma_\pm$ acts only on the second qubit and leaves the first qubit unchanged.

We choose an initially entangled state of the form
\begin{equation}
\ket{\psi_e(0)} = \alpha \ket{01} + \sqrt{1 - \alpha^2} \ket{10},
\end{equation}
whose time evolution under the Lindblad dynamics yields the following density matrix in the computational basis~\cite{Saleem2025}
\begin{equation}
\rho_e(t) =
\begin{pmatrix}
\rho_{11}(t) & 0 & 0 & 0 \\
0 & \rho_{22}(t) & \rho_{23}(t) & 0 \\
0 & \rho_{32}(t) & \rho_{33}(t) & 0 \\
0 & 0 & 0 & 0
\end{pmatrix},
\end{equation}
where the nonzero components are given by
\begin{align}
\rho_{11}(t) &= 1 - e^{-\kappa t}, \nonumber \\
\rho_{22}(t) &= \tfrac{e^{-\kappa t}}{2} \left[1 - (1 - 2\alpha^2)\cos(2gt)\right], \nonumber \\
\rho_{33}(t) &= \tfrac{e^{-\kappa t}}{2} \left[1 + (1 - 2\alpha^2)\cos(2gt)\right], \nonumber \\
\rho_{23}(t) &= \tfrac{e^{-\kappa t}}{2} \left[2\alpha \sqrt{1 - \alpha^2} + i(2\alpha^2 - 1)\sin(2gt)\right], \nonumber \\
\rho_{32}(t) &= \rho_{23}^*(t). \label{eq:JC_components}
\end{align}

Using the modular reflection of the density matrix \( \rho_e \) under \( J_{\mathrm{QIT}}\),
we construct the modular product matrix as
\begin{equation}
R = \rho_e(t)\, ( J_{\mathrm{QIT}} \; \rho_e(t) \; J_{\mathrm{QIT}}\bigr) =
\begin{pmatrix}
R_{11} & 0 & 0 & 0 \\
0 & R_{22} & R_{23} & 0 \\
0 & R_{32} & R_{33} & 0 \\
0 & 0 & 0 & R_{44}
\end{pmatrix}, \label{Rmatrix}
\end{equation}
Let \( \lambda_1 \geq \lambda_2 \) denote the nonzero eigenvalues of \( R \). The concurrence is given by
\begin{equation}
C_e(g,t) = \max \left\{ 0, \sqrt{\lambda_1} - \sqrt{\lambda_2} \right\}.
\end{equation}
where a direct computation yields
\begin{equation}
\begin{array}{c}
\lambda_{1,2} = \dfrac{e^{-2\kappa t}}{4} \left( 1 \pm \sqrt{1 - (1 - 2\alpha^2)^2 \cos^2(2gt)} \right)^2, \\\\
C_e(g,t) = e^{-\kappa t} \sqrt{1 - (1 - 2\alpha^2)^2 \cos^2(2gt)}
\end{array}
\end{equation}
as in \cite{Saleem2025}.
Although the Lindblad dynamics introduce population decay into the ground state \(\ket{00}\), this contribution has no effect on the concurrence at any time.
 Amplitude damping causes population leakage from \(\ket{01}\) and \(\ket{10}\) into \(\ket{00}\), leading to \(\rho_{11}(t) = 1 - e^{-\kappa t}\). However, the concurrence depends only on the nonzero eigenvalues of the matrix \( R = \rho \rho_J \). Under the action of modular conjugation, the state \(\ket{00}\) maps to \(\ket{11}\), which remains unpopulated throughout the evolution. Consequently, the \(\rho_{11}(t)\) term does not contribute to \(R\) and plays no role in the computation of concurrence. This decoupling of the ground state from the entanglement dynamics reflects a structural separation between the excitation-preserving sector and the dissipative decay channel, and ensures that the concurrence remains determined solely by coherence and populations within the \(\{\ket{01}, \ket{10}\}\) subspace. Let \( \beta := 1 - 2\alpha^2 \). The modular curvature of entanglement is then given by
\begin{equation}
\begin{array}{c}
\mathcal{K}_E(g,t)= \dfrac{\partial^2 C(\rho(g,t))}{\partial g^2}\\\\= -4 t^2 e^{-\kappa t} \; \beta^2 \; \dfrac{1 - 2\cos^2(2gt) + \beta^2 \cos^4(2gt)}{[1 - \beta^2 \cos^2(2gt)]^{3/2}}.
\end{array}
\end{equation}

\bigskip

We next compute the quantum Fisher information (QFI) \( F(g, t) \) for the dissipative state \( \rho_e(g, t) \).
The QFI measure quantifies the infinitesimal distinguishability between nearby quantum states and sets a fundamental limit on the precision of parameter estimation in quantum metrology. Given a one-parameter family of density matrices \( \rho(\theta) \), the QFI is defined via the symmetric logarithmic derivative (SLD) \( L_\theta \), which satisfies
\begin{equation}
\frac{d\rho(\theta)}{d\theta} = \frac{1}{2} \left( \rho(\theta) L_\theta + L_\theta \rho(\theta) \right).
\end{equation}
The quantum Fisher information is then given by
\begin{equation}
F_Q(\theta) = \mathrm{Tr} \left[ \rho(\theta) L_\theta^2 \right].
\end{equation}
When \( \rho(\theta) \) is diagonalizable as \( \rho(\theta) = \sum_i \lambda_i(\theta) |\phi_i(\theta)\rangle \langle \phi_i(\theta)| \), the QFI admits an explicit eigenvalue expansion:
\begin{equation}
F_Q(\theta) = \sum_{i} \frac{(\partial_\theta \lambda_i)^2}{\lambda_i} + 2 \sum_{i \neq j} \frac{(\lambda_i - \lambda_j)^2}{\lambda_i + \lambda_j} \left| \langle \phi_i | \partial_\theta \phi_j \rangle \right|^2.
\end{equation}
The first term captures the classical Fisher information associated with changes in eigenvalues, while the second term encodes the quantum contribution arising from the parameter dependence of the eigenbasis.

Although \(\rho_e(g,t)\) develops population in \(\ket{00}\) due to amplitude damping, the entanglement is entirely supported in the \(\{\ket{01}, \ket{10}\}\) subspace. Thus, for the purpose of computing concurrence and QFI, \(\rho_e(g,t)\) behaves effectively as a rank-2 mixed state within this excitation-preserving sector.
We therefore compute the quantum Fisher information (QFI) using the standard expression for mixed states, restricted to a two-dimensional subspace:
\begin{equation}
F(g, t) = 2 \sum_{m,n} \frac{\left| \langle \lambda_m | \partial_g \rho_e(g, t) | \lambda_n \rangle \right|^2}{\lambda_m + \lambda_n},
\end{equation}
where \( \{ \lambda_n, |\lambda_n\rangle \} \) denote the eigenvalues and eigenvectors of the reduced density matrix \( \rho_e(g, t) \), and the sum is taken over all indices \( m,n \) such that \( \lambda_m + \lambda_n \ne 0 \). This form of the QFI arises by solving the symmetric logarithmic derivative (SLD) equation for \( \partial_g \rho_e \) and computing the trace of \( \rho_e L_g^2 \) in its instantaneous eigenbasis. It provides an exact expression valid for any full-rank density matrix and compactly encodes both classical and quantum contributions: changes in eigenvalues (spectrum) and changes in eigenvectors (coherence structure) of \( \rho_e(g, t) \). When \( \rho_e \) is restricted to a two-dimensional support, this expression is especially tractable and avoids explicit differentiation of the eigenvalues or eigenvectors themselves. It is equivalent to the more familiar eigenvalue expansion of the QFI above
but repackages all derivatives into matrix elements of \( \partial_g \rho_e \), making it more amenable to numerical evaluation and physical interpretation in the eigenbasis of \( \rho_e \).

Let us now evaluate this expression at the symmetry points \( gt = \frac{(2n+1)\pi}{4} \), where \( \cos(2gt) = 0 \).
 At these points, the matrix elements simplify to
\begin{equation}
\begin{array}{c}
\rho_{22}(t) = \rho_{33}(t) = \frac{e^{-\kappa t}}{2}, \\\\ \mathrm{Re}[\rho_{23}(t)] = \alpha\sqrt{1 - \alpha^2} e^{-\kappa t}, \\\\ \mathrm{Im}[\rho_{23}(t)] = 0.
\end{array}
\end{equation}
As a result, the density matrix \( \rho_e(g, t) \) becomes real and symmetric, with two nonzero eigenvalues,
\begin{equation}
\lambda_{\pm} = e^{-\kappa t} \left( \frac{1}{2} \pm \alpha \sqrt{1 - \alpha^2} \right).
\end{equation}
The only \( g \)-dependence at these points enters through the imaginary part of \( \rho_{23} \), so \( \partial_g \rho_e(g, t) \) is nonzero only in the \( \rho_{23} \) and \( \rho_{32} \) entries, and only through
\begin{equation}
\partial_g \rho_{23}(t) = i(2\alpha^2 - 1)\, t\, \cos(2gt)\, e^{-\kappa t}.
\end{equation}
Since \( \cos(2gt) = 0 \) at these points, it follows that
\begin{equation}
\partial_g \rho_e(g, t) = 0.
\end{equation}
Hence, the first derivative vanishes, and the curvature of entanglement arises solely from the second derivative of \( \rho_e \).

The quantum Fisher information at these points is calculated directly from the eigenvalue-based analytic formula above giving
\begin{equation}
F(g, t) = 4t^2 e^{-\kappa t}(1 - 2\alpha^2)^2.
\end{equation}
The curvature of entanglement at the same symmetry point simplifies as 
\begin{equation}
K_E(g, t) = -4t^2 e^{-\kappa t}(1 - 2\alpha^2)^2.
\end{equation}
Hence, we recover the relation
\begin{equation}
K_E(g, t) = -F(g, t),
\end{equation}
as first established in \cite{Saleem2025}. The negative sign in this identity reflects our convention of defining \( \mathcal{K}_E(g) \) as the curvature of a concave entanglement functional. Since many entanglement monotones, including those expressible as \( f(\rho, J \rho J) \), are concave in the state argument, their second derivative yields a negative value. This sign ensures that \( \mathcal{K}_E \) and the quantum Fisher information \( F(g) \) are equal in magnitude but capture dual operational meanings: the former as a measure of entanglement deformation and the latter as a bound on parameter sensitivity. We emphasize that this minus sign is not fundamental but depends on the convexity properties of the functional used to define \( \mathcal{K}_E \).
The modular curvature quantifies the second-order deviation of the entanglement from modular self-duality under variations in the coupling strength \( g \), and matches the quantum Fisher information at points of exact modular symmetry.

These results offer a concrete physical interpretation of modular curvature in open quantum dynamics by linking it to the structure of modular flow. Given a full-rank state \( \rho_e(t) \), the modular operator \( \Delta = \rho_e(t) \otimes \rho_e(t)^{-1} \) generates a one-parameter automorphism group on the algebra of observables via the modular flow
\begin{equation}
\sigma_s(A) := \Delta^{is} A \Delta^{-is}.
\end{equation}
This defines an intrinsic, state-dependent evolution governed by the modular Hamiltonian 
\begin{equation} H_{\mathrm{mod}}(t) = -\log \rho_e(t) \end{equation} 
distinct from any external unitary dynamics generated by model Hamiltonians.

At points of exact modular self-duality, where \( \rho_e = J \rho_e J \), the state is symmetric under modular reflection, and the modular flow preserves the entangled structure of the system. In this regime, the modular curvature \( \mathcal{K}_E(g,t) \), defined as the second derivative of the modular overlap, coincides with the quantum Fisher information \( F(g,t) \) but with a negative sign, $\mathcal{K}_E(g,t) = -F(g,t).$
This identity signifies that the quantum state's entanglement is maximally stable under infinitesimal deformations in \( g \), and that the modular flow defines a co-moving reference frame in which the state's entanglement remains stationary. This operator-algebraic perspective shows how modular symmetry constrains dynamical entanglement evolution in open quantum systems.

As the state evolves under dissipation and interaction, \( \rho_e(t) \) deviates from its modular reflection \( J \rho_e(t) J \) whereby the modular curvature \( \mathcal{K}_E(g,t) \) quantifies the rate at which this symmetry is broken. Here, the modular curvature measures the system’s dynamical departure from the co-moving reference frame of entanglement. This interpretation links modular geometry to intrinsic quantum dynamics via the operator-algebraic setting.

\subsection{Relative Entropy, Curvature, and Modular Sensitivity}

In quantum information theory, the \emph{relative entropy} between two faithful normal states \( \rho \) and \( \sigma \) is defined as  
\begin{equation}
S(\rho \| \sigma) := \operatorname{Tr}\!\left(\rho \log \rho - \rho \log \sigma\right), \label{eq:rel_entropy}
\end{equation}
and serves as a fundamental measure of statistical distinguishability between quantum states. It plays a central role in quantum thermodynamics, hypothesis testing, and information-theoretic bounds on state conversion~\cite{Petz1,Petz2,Haag1}.  
Equation~\eqref{eq:rel_entropy} is well defined whenever \( \rho \) and \( \sigma \) are density operators on a finite-dimensional Hilbert space or, more generally, normal states on a \emph{type~I} von~Neumann algebra.

\bigskip

For \emph{type~III} von~Neumann algebras—ubiquitous in algebraic quantum field theory—no normal trace exists, and hence the expression~\eqref{eq:rel_entropy} cannot be defined directly.  In this setting, the operator-algebraic framework of Tomita--Takesaki modular theory provides a general, trace-independent formulation of relative entropy.  
Let \( \mathfrak{M} \subset \mathcal{B}(\mathcal{H}) \) denote a von~Neumann algebra of bounded operators acting on a Hilbert space \( \mathcal{H} \), closed in the weak operator topology and containing the identity operator.  The algebra \( \mathfrak{M} \) represents the physical observables associated with a given system or spacetime region.  A \emph{faithful normal state} \( \rho \) on \( \mathfrak{M} \) is represented by a cyclic and separating vector \( |\Omega_\rho\rangle \in \mathcal{H} \) satisfying
\begin{equation}
\rho(A) = \langle \Omega_\rho | A | \Omega_\rho \rangle, \qquad A \in \mathfrak{M}.
\end{equation}
The pair \( (\mathfrak{M},|\Omega_\rho\rangle) \) thus forms the \emph{standard representation} of the algebra.
Following Araki~\cite{Araki}, one associates to two such faithful normal states \( \rho \) and \( \sigma \) the \emph{relative modular operator} \( \Delta_{\sigma,\rho} \), defined by
\begin{equation}
\Delta_{\sigma,\rho} = S_{\sigma,\rho}^\dagger S_{\sigma,\rho},
\end{equation}
where the \emph{relative Tomita operator} \(S_{\sigma,\rho}\) acts on the dense domain
\begin{equation}
\mathcal{D}(S_{\sigma,\rho}) = \{ A|\Omega_\rho\rangle : A \in \mathfrak{M} \},
\end{equation}
according to
\begin{equation}
S_{\sigma,\rho}\, A|\Omega_\rho\rangle = A^\dagger |\Omega_\sigma\rangle.
\end{equation}
Here, \( |\Omega_\sigma\rangle \) is the GNS vector corresponding to the state \( \sigma \), also cyclic and separating for \( \mathfrak{M} \).

\bigskip

The relative modular operator is characterized by the quadratic form
\begin{equation}
\langle \psi | \Delta_{\sigma,\rho} \phi \rangle
  = \langle S_{\sigma,\rho}\psi | S_{\sigma,\rho}\phi \rangle,
  \qquad \forall\, \psi,\phi \in \mathcal{D}(S_{\sigma,\rho}).
\end{equation}
This construction requires neither a trace nor a matrix representation, making it valid for general type~III factors.  With this operator, Araki’s relative entropy is expressed as
\begin{equation}
S(\rho \| \sigma)
  = -\langle \Omega_\rho | \log \Delta_{\sigma,\rho} | \Omega_\rho \rangle, \label{eq:modular_rel_entropy}
\end{equation}
which generalizes the finite-dimensional expression to arbitrary von~Neumann algebras.  In this formulation, the operator \( \log \Delta_{\sigma,\rho} \) plays the role of the \emph{relative modular Hamiltonian} in the Tomita--Takesaki framework~\cite{Araki,Takesaki1970,Haag1}. This expression remains meaningful in type~III algebras, 
where \(\Delta_{\sigma,\rho}\) has a continuous spectrum 
and the trace-based definition is not applicable.

\bigskip

In the special case of \emph{type~I} von~Neumann algebras, the construction reduces to the familiar finite-dimensional form. One obtains the concrete realization
\begin{equation}
\Delta_{\sigma,\rho} = \sigma \otimes \rho^{-1},
\end{equation}
acting on the Hilbert--Schmidt space with GNS vector
\(|\Omega_\rho\rangle = \sum_i \sqrt{\lambda_i}\,|\phi_i\rangle \otimes |\phi_i\rangle\),
where \(\rho = \sum_i \lambda_i |\phi_i\rangle\langle\phi_i|\).
In this case,
Eq.~\eqref{eq:modular_rel_entropy} reproduces the standard relative entropy found in quantum information theory.

\bigskip

The explicit constructions of the Tomita operator \(S_\rho\), modular conjugation \(J_\rho\), and modular operator \(\Delta_\rho\), together with their modular-flow relations, are summarized in Appendix~A.  A systematic definition of ``modular concurrence'' or ``entanglement curvature'' in genuine 
type~III settings would require replacing density-operator expressions by modular pairs of states 
such as $(\rho, J\rho J)$ interpreted through relative modular operators or via the 
\emph{split property}, which provides an effective type~I inclusion between nested local algebras. 
While a full generalization along these lines lies beyond the present scope, 
the modular-theoretic relations shown here form a consistent operator-algebraic foundation 
from which such type~III extensions can be constructed.

\bigskip

Let \( \rho(g) \) denote a smoothly varying family of density operators parametrized by a real coupling parameter \( g \). 
Consider a perturbed version of \( \rho(g) \) under a small completely positive trace-preserving (CPTP) map 
\( \mathcal{E}_\varepsilon \) such that \( \mathcal{E}_0 = \mathrm{id} \). 
The relative entropy between the unperturbed and perturbed states then satisfies the second-order expansion
\begin{equation}
S(\rho(g) \| \mathcal{E}_\varepsilon(\rho(g))) 
   = \frac{\varepsilon^2}{2} F(g) + \mathcal{O}(\varepsilon^3), 
   \label{eq:second_order_rel_entropy}
\end{equation}
or equivalently,
\begin{equation}
\frac{d^2}{d\varepsilon^2} 
   S(\rho(g) \| \rho(g+\varepsilon))\Big|_{\varepsilon=0} = F(g),
   \label{eq:relative_entropy_expansion}
\end{equation}
where \( F(g) \) denotes the quantum Fisher information associated with the family \( \rho(g) \).
This second-order behavior reveals that \( F(g) \) quantifies the local curvature of the quantum state manifold with respect to the relative entropy. 
Fawzi and Renner~\cite{FawziRenner2015} rigorously established that small relative entropy implies the existence of an almost optimal recovery channel—specifically the Petz map—while Sutter, Tomamichel, and Berta~\cite{SutterTomamichelBerta2017} obtained tight second-order expansions quantifying the degree of irreversibility under noise.
To leading order, a small value of \(F(g)\) indicates that the relative entropy between 
\(\rho(g)\) and \(\rho(g+\varepsilon)\) grows slowly with \(\varepsilon\),
signifying that the state remains nearly reversible under the relevant CPTP channel and is well approximated by the Petz recovery map.
Conversely, a large \(F(g)\) signals that reversibility breaks down sharply—the relative entropy grows rapidly with \(\varepsilon\),
and the state exhibits strong sensitivity to perturbations.

\bigskip

This geometric curvature has an operator-algebraic counterpart in the 
\emph{entanglement curvature} \( \mathcal{K}_E(g) \),
where the modular conjugation operator \(J\) encodes the same local sensitivity captured by the quantum Fisher information, 
but expressed entirely within the Tomita--Takesaki algebraic framework.
Just as relative entropy is defined through the relative modular operator \( \Delta_{\sigma,\rho} \),
the curvature \( \mathcal{K}_E(g) \) reflects second-order variations of modular structure and hence the ``modular sensitivity'' of the entanglement geometry.

\bigskip

While the quantum Fisher information quantifies local statistical distinguishability and optimal parameter estimation,
the modular curvature measures the responsiveness of the entanglement structure under modular conjugation.
At \emph{self-duality points} satisfying \( \rho = J\rho J \),
these two notions coincide,
yielding \( \mathcal{K}_E(g) = -F(g) \);
the equality provides a precise modular signature of reversibility.
Away from self-duality, however, the two curvatures diverge:
\( \mathcal{K}_E(g) \) no longer tracks \(F(g)\) but instead measures the degree of modular asymmetry and the rate at which the state departs from the fixed-point condition 
\( \rho(g) = J\rho(g)J \).
Even in regimes where \(F(g)\) loses its operational meaning—due to singularities, rank deficiency, or non-Markovian dynamics—the modular curvature 
\( \mathcal{K}_E(g) \) continues to furnish a robust geometric indicator of entanglement sensitivity.

\section{Conclusion and Future Research}

In summary, we have presented a modular operator framework for analyzing the curvature of quantum entanglement, grounded in the Tomita–Takesaki theory and modular conjugation. By explicitly constructing entanglement functionals of the form \( f(\rho, J \rho J) \), we have demonstrated how the deviation between a quantum state and its modular reflection encodes geometric sensitivity, informational asymmetry, and entanglement structure. We established a direct correspondence between this curvature and quantum Fisher information at points of modular self-duality, unifying differential geometric and operator-algebraic perspectives on quantum information. This framework not only captures the second-order response of entangled states under parameter variations but also reveals the underlying modular symmetries that govern their evolution, offering a new lens for understanding entanglement dynamics in both discrete and continuous quantum systems.

One may now propose a broader class of entanglement functionals constructed from pairs \( (\rho, J \rho J) \) of the form
\begin{equation}
M(\rho) = f(\rho, J \rho J),
\end{equation}
which are invariant under modular reflection at points of exact modular self-duality where \( \rho = J \rho J \). Functionals of this type vanish when \( \rho \) is orthogonal to its modular image and attain their maximum value when modular symmetry is exact

A general representative of this class can be expressed as
\begin{equation}
M_f(\rho) = \operatorname{Tr}\left[ f(\rho^{1/2} J \rho J \rho^{1/2}) \right],
\end{equation}
where \( f \) is a scalar function on positive operators, such as \( f(x) = x^\alpha \), \( f(x) = -x \log x \), or more generally, an operator monotone function. We assume \( \rho \in \mathcal{T}_1(\mathcal{H}) \) is a positive trace-class operator and that \( f: \mathbb{R}_+ \to \mathbb{R} \) admits a holomorphic functional calculus on the spectrum of \( \rho^{1/2} (J \rho J) \rho^{1/2} \). Under these conditions, the operator \( f(\rho^{1/2} (J \rho J) \rho^{1/2}) \) is trace-class, ensuring that \( M_f(\rho) \) is finite.

These functionals are closely related to sandwiched R\'{e}nyi divergences and Petz--R\'{e}nyi relative entropies, which take the form
\begin{equation}
S_\alpha(\rho \| J \rho J) = \frac{1}{\alpha - 1} \log \operatorname{Tr}\left[ \rho^\alpha (J \rho J)^{1 - \alpha} \right].
\end{equation}
In particular, the Uhlmann fidelity between \( \rho \) and its modular reflection \( J \rho J \) is given by
\begin{equation}
F(\rho, J \rho J) = \left( \operatorname{Tr} \left[ \sqrt{ \sqrt{\rho} J \rho J \sqrt{\rho} } \right] \right)^2,
\end{equation}
and serves as a distance-like measure of entanglement asymmetry. Deviation of this quantity from unity directly quantifies the degree of modular symmetry breaking.

Modular frame curvature emerges naturally by considering the second derivative of such functionals along a one-parameter family \( \rho(g) \).
While we conjecture that entanglement monotones of the form \( M(\rho) = f(\rho, J \rho J) \) naturally induce curvature metrics that coincide with the quantum Fisher information, the correspondence may depend on the smoothness and differentiability of \( f \), as well as on spectral regularity of the family \( \rho(g) \). A precise classification of such functionals, and the parameter regimes under which they yield monotone metrics, remains a subject for future investigation. 

Taken together, these findings suggest that modular operator theory offers a promising framework for relating entanglement structure, curvature, and recoverability within quantum information theory. We expect that the operator-algebraic techniques introduced here extend beyond the specific models studied, providing a modular toolkit for analyzing entanglement dynamics, quantum channel reversibility, and information geometry. Potential applications include quantum metrology~\cite{Giovannetti2011}, quantum resource theories~\cite{Chitambar2019}, and the structure of entanglement in quantum field theory~\cite{Witt2018} and holography~\cite{deBoer2025}.
 Moreover, as noted in Sec.~V(B), the connection to Araki’s relative modular operator \( \Delta_{\sigma,\rho} \) and relative entropy provides a pathway toward generalizing the present framework to type~III von~Neumann algebras, where density-matrix representations are unavailable. Within algebraic quantum field theory, such generalizations would naturally invoke the split property to realize approximate type~I factorizations and to characterize modular sensitivity in local algebras. These extensions connecting modular concurrence and entanglement curvature to the operator-algebraic foundations of field theory will be explored in future work.

\section{Appendix}

See \cite{Takesaki1970, Witt2022, Witt2018, Guido2011, Bratteli1997, Haag1, Takesaki1} for extensive reviews of the topics outlined below.

\subsection{Tomita–Takesaki Modular Operators}

 Consider a  $\bm{C^*}$-algebra $\mathcal{B(H)} =\{A\}$ of bounded linear operators on a Hilbert space, $ A: \mathcal{H} \rightarrow \mathcal{H}$. Let $\mathcal{C}$ be a subset of $\mathcal{B(H)}$. An operator $A \in \mathcal{B(H)} $ belongs to the commutant $\mathcal{C}'$ of the set $\mathcal{C}$ $\iff$   $AC =CA, \,\,\, \forall C \in \mathcal{C}$. A von Neumann algebra $\mathcal{A}$ is a unital $\bm{C^*}$-subalgebra of $\mathcal{B(H)}$ such that $\mathcal{A}'' = \mathcal{A}$. Von Neumann proved that this bicommutant definition is equivalent to the algebra $\mathcal{A}$ being closed with respect to the weak topology on $\mathcal{B(H)}$. A von Neumann algebra in standard form is one where there exists an element $| \Omega \rangle \in \mathcal{H}$ which is both cyclic (operating on $| \Omega \rangle $ with elements in  $\mathcal{A}$ can generate a space dense in $\mathcal{H}$) and separating (if $A | \Omega \rangle = 0$, then $A=0$).

\bigskip

A von Neumann algebra in standard form is one where there exists an element \( | \Omega \rangle \in \mathcal{H} \) which is both cyclic (operating on \( | \Omega \rangle \) with elements in \( \mathcal{A} \) can generate a space dense in \( \mathcal{H} \)) and separating (if \( A | \Omega \rangle = 0 \), then \( A = 0 \)). Such a vector \( |\Omega\rangle \) arises naturally from the Gelfand--Naimark--Segal (GNS) construction associated to any faithful normal state \( \omega \) on \( \mathcal{A} \). In this framework, one constructs a Hilbert space \( \mathcal{H}_\omega \), a representation \( \pi_\omega: \mathcal{A} \to \mathcal{B}(\mathcal{H}_\omega) \), and a cyclic vector \( |\Omega_\omega\rangle \in \mathcal{H}_\omega \) such that
\begin{equation}
\omega(A) = \langle \Omega_\omega | \pi_\omega(A) | \Omega_\omega \rangle \quad \text{for all } A \in \mathcal{A}.
\end{equation}
If the state \( \omega \) is faithful, then \( |\Omega_\omega\rangle \) is not only cyclic but also separating for \( \pi_\omega(\mathcal{A}) \). Thus, the GNS construction ensures that \( \mathcal{A} \) is in standard form, which is the setting required for Tomita--Takesaki modular theory.

 Let $S:\mathcal{H} \rightarrow \mathcal{H}$ be an anti-linear, densely defined operator given by $S A  | \Omega \rangle = A^* | \Omega \rangle $. Let the closure of $S$ have a polar decomposition given by $S=J \Delta ^{\frac{1}{2}} =\Delta^{-\frac{1}{2}} J$, where $J$ is called the modular conjugation operator and $\Delta$ is called the modular operator. $J$ is anti-linear and anti-unitary whereas $\Delta$ is self-adjoint and positive. Furthermore, the following relations hold:

1. $J \Delta ^{\frac{1}{2}} J = \Delta ^{-\frac{1}{2}}$

2. $ J^2 =I, \,\,\, J^{*} = J$

3. $ J |\Omega \rangle  = |\Omega \rangle  $

4. $ J \mathcal{A} J = \mathcal{A}' $

5. $ \Delta = S^{*} S$

6. $ \Delta |\Omega \rangle  = |\Omega \rangle $

\noindent The modular automorphism group is then defined by
\begin{equation}
\sigma_t(A) = \Delta^{it} A \Delta^{-it}, \quad \forall t \in \mathbb{R}, \; A \in \mathcal{A}.
\end{equation}
which is referred to as the modular flow.
The modular flow $\sigma_t$ plays a central role in understanding time evolution in algebraic quantum field theory, thermality of vacuum states (e.g., Bisognano–Wichmann theorem), and the geometry of quantum entanglement via Tomita–Takesaki theory.

\subsection{Explicit Realizations of the Modular Conjugation Operator}

\subsubsection{Finite Dimensional}
Let \( \mathcal{A} \subset \mathcal{B}(\mathcal{H}) \) be a von Neumann algebra with a cyclic and separating vector \( |\Omega\rangle \), and define the GNS purification (in the doubled Hilbert space $\mathcal{H} \otimes \mathcal{H}$) using the spectral decomposition of a full-rank state \( \rho = \sum_i \lambda_i |\phi_i\rangle\langle \phi_i| \) with orthonormal eigenbasis \( \{ \ket{\phi_i} \} \) as
\begin{equation}
|\Omega\rangle = \sum_i \sqrt{\lambda_i} |\phi_i\rangle \otimes |\phi_i\rangle .
\end{equation}
The modular conjugation operator \( J \) acts anti-linearly on the dense set of vectors in \( \mathcal{H} \otimes \mathcal{H} \), corresponding to the algebra \( \mathcal{A} = \mathcal{B}(\mathcal{H}) \otimes I \) and its commutant \( \mathcal{A}' = I \otimes \mathcal{B}(\mathcal{H})\) 
\begin{equation}
\sum_{i,j} c_{ij} |\phi_i\rangle \otimes |\phi_j\rangle
\end{equation}
as follows
\begin{equation}
J\left( \sum_{i,j} c_{ij} |\phi_i\rangle \otimes |\phi_j\rangle \right)
= \sum_{i,j} \overline{c_{ji}} |\phi_j\rangle \otimes |\phi_i\rangle .
\end{equation}

For any operator \( A \in \mathcal{A} \), acting as \( A \otimes I \), we compute:
\begin{equation}
\begin{array}{c}
J (A \otimes I) J \left( \sum_{i,j} c_{ij} |\phi_i\rangle \otimes |\phi_j\rangle \right) \\[6pt] 
=J (A \otimes I)\left(\sum_{i,j} \overline{c_{ji}} |\phi_j\rangle \otimes |\phi_i\rangle\right)
\\[6pt] 
= J\left(\sum_{i,j,k} \overline{c_{ji}} \, a_{kj} \, |\phi_k\rangle \otimes |\phi_i\rangle \right) \\[6pt] 
= \sum_{i,j,k} c_{ij} \, \overline{a_{jk}} \, |\phi_i\rangle \otimes |\phi_k\rangle ,
\end{array}
\end{equation}
where we have used \( A|\phi_j\rangle = \sum_k a_{kj} |\phi_k\rangle \). Since  \(A^\dagger |\phi_j\rangle = \sum_k \overline{a_{jk}} |\phi_k\rangle\), we have  
\begin{equation}
\begin{array}{c}
(I \otimes A^\dagger) \left( \sum_{i,j} c_{ij} |\phi_i\rangle \otimes |\phi_j\rangle \right) \\[6pt] = \sum_{i,j} c_{ij} |\phi_i\rangle \otimes \left( \sum_k \overline{a_{jk}} |\phi_k\rangle \right).
\end{array}
\end{equation}
and hence,
\begin{equation}
J (A \otimes I) J = I \otimes A^\dagger
\end{equation}
where \( A^\dagger \) is the Hermitian adjoint. Therefore,
\begin{equation}
J \mathcal{A} J = \mathcal{A}'
\end{equation}
as expected from Tomita--Takesaki theory. 

Let
\begin{equation}
\begin{array}{c}
\ket{\psi} = \sum_{i,j} c_{ij} \ket{\phi_i} \otimes \ket{\phi_j} \\[6pt] 
\ket{\xi} = \sum_{l,m} d_{lm} \ket{\phi_l} \otimes \ket{\phi_m}
\end{array}
\end{equation}
Then the inner product is
\begin{equation}
\begin{array}{c}
\braket{\psi | \xi}
= \sum_{i,j,l,m} \overline{c_{ji}} d_{lm}
\braket{\phi_i|\phi_l} \braket{\phi_j|\phi_m} \\[6pt] 
= \sum_{i,j} \overline{c_{ji}} d_{ij}
\end{array}
\end{equation}
Apply \( J \):
\begin{equation}
\begin{array}{c}
J \ket{\psi} = \sum_{i,j} \overline{c_{ji}} \ket{\phi_j} \otimes \ket{\phi_i} \\[6pt] 
J \ket{\xi} = \sum_{l,m} \overline{d_{ml}} \ket{\phi_m} \otimes \ket{\phi_l}
\end{array}
\end{equation}
Now compute the inner product \( \braket{J \xi | J \psi} \),
\begin{equation}
\begin{array}{c}
\braket{J \xi | J \psi}
= \sum_{i,j,l,m} d_{lm}\overline{c_{ji}}
\braket{\phi_m | \phi_j} \braket{\phi_l | \phi_i} \\\\
= \sum_{i,j} \overline{c_{ji}} d_{ij}= \overline{\braket{\xi | \psi}}=\braket{\psi | \xi}
\end{array}
\end{equation}
Since complex conjugation is anti-linear and the conjugation reverses the tensor structure, this demonstrates that \( J \) is anti-linear and norm-preserving.
This confirms that \( J \) is anti-unitary,
\begin{equation}
\braket{J \xi | J \psi} = \overline{\braket{\xi | \psi}}
\end{equation}

The modular operator \( \Delta \) is defined by
\begin{equation}
\Delta = \rho \otimes \rho^{-1}
= \sum_{i,j} \frac{\lambda_i}{\lambda_j}\, 
\ket{\phi_i}\bra{\phi_i} \otimes \ket{\phi_j}\bra{\phi_j} .
\end{equation}
and leaves \( |\Omega\rangle \) invariant, 
\begin{equation}
\begin{array}{c}
\Delta |\Omega\rangle = (\rho \otimes \rho^{-1}) \left( \sum_i \sqrt{\lambda_i} |\phi_i\rangle \otimes |\phi_i\rangle \right)
\\\\ = \sum_{i,j,k} \frac{\lambda_i}{\lambda_j} \sqrt{\lambda_k} \left( |\phi_i\rangle\langle \phi_i| \otimes |\phi_j\rangle\langle \phi_j| \right) \left( |\phi_k\rangle \otimes |\phi_k\rangle \right) \\\\
= \sum_{i,j,k} \frac{\lambda_i}{\lambda_j} \sqrt{\lambda_k} \delta_{ik} \delta_{jk} |\phi_i\rangle \otimes |\phi_j\rangle \\\\
= \sum_i \frac{\lambda_i}{\lambda_i} \sqrt{\lambda_i} |\phi_i\rangle \otimes |\phi_i\rangle 
= |\Omega\rangle
\end{array}
\end{equation}

\medskip

We now define the Tomita operator \( S \) via its polar decomposition:
\begin{equation}
S = J \Delta^{1/2} .
\end{equation}
and verify the defining property $S A|\Omega\rangle = A^*|\Omega\rangle $.
For any \( A \in \mathcal{B}(\mathcal{H}) \). 
\begin{equation}
\begin{array}{c}
S (A \otimes I) |\Omega\rangle = J \Delta^{1/2} (A \otimes I) |\Omega\rangle \\[6pt] 
= J \Delta^{1/2}(A \otimes I) \left(\sum_j \sqrt{\lambda_j} |\phi_j\rangle \otimes |\phi_j\rangle \right) \\[6pt] =
 J \Delta^{1/2} \left(\sum_{j,k} \sqrt{\lambda_j} a_{kj}|\phi_k\rangle \otimes |\phi_j\rangle \right) \\[6pt] 
= J \left(\sum_{j,k} \sqrt{\lambda_k} a_{kj}|\phi_k\rangle \otimes |\phi_j\rangle \right)  \\[6pt] 
= \sum_{j,k} \sqrt{\lambda_j} \overline{a_{jk}}|\phi_j\rangle \otimes |\phi_k\rangle  \\[6pt] 
= (A^* \otimes I) \sum_j \sqrt{\lambda_j} |\phi_j\rangle \otimes |\phi_j\rangle 
= (A^* \otimes I) |\Omega\rangle
\end{array}
\end{equation}

\subsubsection{Qubits}

We consider the special case of the modular conjugation operator \(J_{AB}\) acting on product \textit{qubit states} in \(\mathcal{H}_A \otimes \mathcal{H}_B \cong \mathbb{C}^2 \otimes \mathbb{C}^2\), defined in the computational basis as
\begin{equation}
J_{AB}\left[
\begin{pmatrix}
\alpha \\
\beta
\end{pmatrix} \otimes
\begin{pmatrix}
\gamma \\
\delta
\end{pmatrix}  \right]
=
\begin{pmatrix}
\bar{\delta} \\
\bar{\gamma}
\end{pmatrix} \otimes
\begin{pmatrix}
\bar{\beta} \\
\bar{\alpha}
\end{pmatrix} .
\end{equation}
This map performs complex conjugation and exchanges the tensor factors while reversing the spin basis order in each qubit. It is valid as long as one is working in the computational qubit basis which is typically the case.

In quantum information theory (QIT), the spin-flip operator used in Wootters' definition of concurrence is
\begin{equation}
J_{\mathrm{QIT}} = (\sigma_y \otimes \sigma_y) K,
\end{equation}
where
\begin{equation}
\sigma_y = \begin{pmatrix}
0 & -i \\
i & 0
\end{pmatrix}
\end{equation}
and $K$ denotes complex conjugation in the computational basis.
Let us compute the action of \(J_{\mathrm{QIT}}\) on a generic product state
\begin{equation}
\ket{\psi} =
\begin{pmatrix}
\alpha \\
\beta
\end{pmatrix} \otimes
\begin{pmatrix}
\gamma \\
\delta
\end{pmatrix}.
\end{equation}
After complex conjugation followed by \(\sigma_y \otimes \sigma_y\), we obtain:
\begin{equation}
J_{\mathrm{QIT}} \ket{\psi}
=
\begin{pmatrix}
\bar{\beta} \\
-\bar{\alpha}
\end{pmatrix}
\otimes
\begin{pmatrix}
-\bar{\delta} \\
\bar{\gamma}
\end{pmatrix}.
\end{equation}
Clearly, the two results differ by a SWAP and additional local sign flips. This is captured precisely by local Pauli-\(\sigma_z\) operators 
\begin{equation}
\sigma_z = \begin{pmatrix}
1 & 0 \\
0 & -1
\end{pmatrix}
\end{equation}
such that (up to global phase)
\begin{equation}
J_{AB} = (\sigma_z \otimes \sigma_z)\, \mathrm{SWAP} \circ J_{\mathrm{QIT}},
\end{equation}
where
\begin{equation}
 \mathrm{SWAP} \begin{pmatrix}
\alpha \\
\beta
\end{pmatrix} \otimes
\begin{pmatrix}
\gamma \\
\delta
\end{pmatrix}= \begin{pmatrix}
\gamma \\
\delta
\end{pmatrix} \otimes  \begin{pmatrix}
\alpha \\
\beta
\end{pmatrix}
\end{equation}
This is a valid modular conjugaton operator as long as one is in the Bell State qubit basis for \(\mathcal{H}_A \otimes \mathcal{H}_B \cong \mathbb{C}^2 \otimes \mathbb{C}^2\). Thus, although \( J_{AB} \) and \( J_{\mathrm{QIT}} \) differ by a local unitary conjugation and SWAP, they define equivalent notions of modular conjugation when restricted to the Bell basis, which is the relevant sector for two-qubit entanglement.

\subsection{Derivation of the Modular Expectation Value \( 2|X| \)}

We compute the modular expectation value
\begin{equation}
\langle \Psi_f | J_{AB} \otimes \mathbb{I}_\phi | \Psi_f \rangle,
\end{equation}
to second order in the interaction strength \( \lambda \). Let the initial state be
\begin{equation}
|\Psi_i\rangle = |0\rangle_A \otimes |0\rangle_B \otimes |0\rangle_\phi.
\end{equation}

The time-evolved final state is expanded perturbatively as
\begin{equation}
|\Psi_f\rangle = |\Psi^{(0)}_f\rangle + \lambda |\Psi^{(1)}_f\rangle + \lambda^2 |\Psi^{(2)}_f\rangle + \mathcal{O}(\lambda^3),
\end{equation}
where \( |\Psi^{(0)}_f\rangle = |\Psi_i\rangle \).
The first-order correction is
\begin{equation}
\begin{array}{c}
|\Psi^{(1)}_f\rangle =  
-i \displaystyle\int dt\, \biggl[ \eta_A(t) e^{i \omega_A \tau_A(t)} |1\rangle_A \otimes |0\rangle_B \otimes \phi(x_A(t))|0\rangle  \\[6pt] 
 + \eta_B(t) e^{i \omega_B \tau_B(t)} |0\rangle_A \otimes |1\rangle_B \otimes \phi(x_B(t))|0\rangle \biggr].
\end{array}
\end{equation}
where we have defined $\eta(t)=\chi(x(\tau))\frac{d\tau}{dt}$.
The relevant second-order term for \( J_{AB} \otimes \mathbb{I}_\phi \) comes from the component of \( |\Psi^{(2)}_f\rangle \) in the \( |1\rangle_A \otimes |1\rangle_B \) subspace (\cite{Smith2019})
\begin{equation}
\begin{array}{c}
|\Psi^{(2)}_{f}\rangle \supset -\frac{1}{2} \displaystyle\int dt\, dt'\, \eta_A(t) \eta_B(t') e^{i (\omega_A \tau_A(t) + \omega_B \tau_B(t'))} \\[6pt] 
\times |1\rangle_A \otimes |1\rangle_B \\[6pt] \otimes \biggl\{\mathcal{T}\left[ \phi(x_B(t')) \phi(x_A(t)) \right] +\mathcal{T}\left[ \phi(x_A(t)) \phi(x_B(t')) \right]\biggr\}|0\rangle.
\end{array}
\end{equation}
Now compute
\begin{equation}
\begin{array}{c}
\langle \Psi_f | J_{AB} \otimes \mathbb{I}_\phi | \Psi_f \rangle \\[6pt] = \displaystyle\sum_{n,m=0}^1 \lambda^{n+m} \langle \Psi_f^{(n)} | J_{AB} \otimes \mathbb{I}_\phi | \Psi_f^{(m)} \rangle + \mathcal{O}(\lambda^3).
\end{array}
\end{equation}
Only cross-terms that map between single-excitation and double-excitation subspaces contribute. The first non-zero term at order \( \lambda^2 \) is:
\begin{equation}
\langle \Psi_f^{(1)} | J_{AB} \otimes \mathbb{I}_\phi | \Psi_f^{(1)} \rangle.
\end{equation}
The second non zero term of order \( \lambda^2 \)is
\begin{equation}
\langle \Psi_f^{(0)} | J_{AB} \otimes \mathbb{I}_\phi | \Psi_f^{(2)} \rangle +\langle \Psi_f^{(2)} | J_{AB} \otimes \mathbb{I}_\phi | \Psi_f^{(0)} \rangle.
\end{equation}
All other terms such as \( \langle \Psi^{(0)}_f | J_{AB} \otimes \mathbb{I}_\phi | \Psi^{(0)}_f \rangle \) vanish due to orthogonality, and \( \langle \Psi^{(2)}_f | J_{AB}\otimes\mathbb{I}_{\phi} | \Psi^{(2)}_f \rangle \sim \lambda^4 \).
First,
\begin{equation}
\begin{array}{c}
(J_{AB} \otimes \mathbb{I}_\phi) \ket{\Psi_f^{(1)}}
\\[6pt] = +i \displaystyle\int dt \biggl[ \eta_A(t) e^{-i\omega_A \tau_A(t)} \ket{0}_A \ket{1}_B \phi[x_A(t)] \ket{0}  \\[6pt] 
 +\ \eta_B(t) e^{-i\omega_B \tau_B(t)} \ket{1}_A \ket{0}_B \phi[x_B(t)] \ket{0} \biggr]
\end{array}
\end{equation}
and
\begin{equation}
\begin{array}{c}
\bra{\Psi_f^{(1)}} = i \displaystyle\int dt' \biggl[ \eta_A(t') e^{-i\omega_A \tau_A(t')} \bra{1}_A \bra{0}_B \bra{0}\phi[x_A(t')]  \\[6pt] 
 +\ \eta_B(t') e^{-i\omega_B \tau_B(t')} \bra{0}_A \bra{1}_B \bra{0}\phi[x_B(t')]\biggr]
\end{array}
\end{equation}
gives the expectation value
\begin{equation}
\begin{array}{c}
\lambda^2 \braket{\Psi_f^{(1)} | J_{AB} \otimes \mathbb{I}_\phi | \Psi_f^{(1)}}
\\[6pt]= (i)(+i)\lambda^2  \int dt' dt \biggl[ \eta_A(t) \eta_B(t') e^{-i[\omega_A \tau_A(t) + \omega_B \tau_B(t')]}  \\[6pt] 
\times \bra{0}  \phi[x_B(t')] \phi[x_A(t)] \ket{0}  \\[6pt] 
 +\ \eta_B(t) \eta_A(t') e^{-i[\omega_A \tau_A(t') + \omega_B \tau_B(t)]}  \\[6pt] \times \bra{0}  \phi[x_A(t')] \phi[x_B(t)] \ket{0} \biggr]
\end{array}
\end{equation}
Using the Wightman function $W(x, x') := \braket{0 | \phi(x) \phi(x') | 0}$ the above becomes
\begin{equation}
\begin{array}{c}
\lambda^2 \braket{\Psi_f^{(1)} | J_{AB} \otimes \mathbb{I}_\phi | \Psi_f^{(1)}} 
= -\lambda^2 \displaystyle\int dt\, dt' \\[6pt] \biggl[
\eta_A(t) \eta_B(t')\, e^{-i\left[\omega_A \tau_A(t) + \omega_B \tau_B(t')\right]}
W\bigl(x_B(t'), x_A(t)\bigr)  \\[6pt]
+ \eta_B(t) \eta_A(t')\, e^{-i\left[\omega_A \tau_A(t') + \omega_B \tau_B(t)\right]}
W\bigl(x_A(t'), x_B(t)\bigr)
\biggr] \\[6pt] \equiv X
\end{array}
\end{equation}
(see \cite{Smith2019}).

Finally, using
\begin{equation}
\begin{array}{c}
-\frac{1}{2}(J_{AB} \otimes \mathbb{I}_\phi)  \displaystyle\int dt\, dt'\, \eta_A(t) \eta_B(t') e^{i (\omega_A \tau_A(t) + \omega_B \tau_B(t'))} \\[6pt] 
\times |1\rangle_A \otimes |1\rangle_B \\[6pt] \otimes \biggl\{\mathcal{T}\left[ \phi(x_B(t')) \phi(x_A(t)) \right] +\mathcal{T}\left[ \phi(x_A(t)) \phi(x_B(t')) \right]\biggr\}|0\rangle \\[6pt] 
=-\frac{1}{2}\displaystyle\int dt\, dt'\, \eta_A(t) \eta_B(t') e^{-i (\omega_A \tau_A(t) + \omega_B \tau_B(t'))} \\[6pt] 
\times |0\rangle_A \otimes |0\rangle_B \\[6pt] \otimes \biggl\{\mathcal{T}\left[ \phi(x_B(t')) \phi(x_A(t)) \right] +\mathcal{T}\left[ \phi(x_A(t)) \phi(x_B(t')) \right]\biggr\}|0\rangle
\end{array}
\end{equation}
it is straightforward to show that 
$\langle \Psi_f^{(0)} | J_{AB} \otimes \mathbb{I}_\phi | \Psi_f^{(2)} \rangle +\langle \Psi_f^{(2)} | J_{AB} \otimes \mathbb{I}_\phi | \Psi_f^{(0)} \rangle = X$.
Thus we obtain
$\left| \langle \Psi_f | J_{AB} \otimes \mathbb{I}_\phi | \Psi_f \rangle \right|
= 2 |X| $
up to $\mathcal{O}(\lambda^4)$ which proves equation (\ref{keyResult}).

\begin{acknowledgments}
The author thanks an anonymous referee for several insightful comments that improved the clarity and scope of this work.
\end{acknowledgments}

\end{document}